\documentclass[12pt]{spieman}  % 12pt font required by SPIE;
\usepackage{amsmath,amsfonts,amssymb}
\usepackage{graphicx}
\usepackage{setspace}
\usepackage{tocloft}
\usepackage{array}
\usepackage{multirow}
\usepackage{upgreek}
\usepackage{booktabs}
\usepackage{paralist}

\newcommand{\micron}[0]{$~\upmu$m }
\newcommand{\btheta}{\boldsymbol{\theta}}
\newcommand{\bt}{\mathbf{t}}

\title{Lightweight starshade position sensing with convolutional neural networks and simulation-based inference}

\author[a]{Andrew Chen}
\author[b]{Anthony Harness}
\author[c,d,*]{Peter Melchior}
\affil[a]{Princeton University, Department of Computer Science, Princeton, New Jersey}
\affil[b]{Princeton University, Department of Mechanical \& Aerospace Engineering, Princeton, New Jersey}
\affil[c]{Princeton University, Department of Astrophysical Sciences, Princeton, New Jersey}
\affil[d]{Princeton University, Center for Statistics and Machine Learning, Princeton, New Jersey}

\cftpagenumbersoff{figure}
\cftpagenumbersoff{table} 

\begin{document} 
\maketitle

\begin{abstract}
Starshades are a leading technology to enable the direct detection and spectroscopic characterization of Earth-like exoplanets.
To keep the starshade and telescope aligned over large separations, reliable sensing of the peak of the diffracted light of the occluded star is required.
Current techniques rely on image matching or model fitting, both of which put substantial computational burdens on resource-limited spacecraft computers.
We present a lightweight image processing method based on a convolutional neural network paired with a simulation-based inference technique to estimate the position of the spot of Arago and its uncertainty.
The method achieves an accuracy of a few centimeters across the entire pupil plane, while only requiring 1.6 MB in stored data structures and 5.3 MFLOPs (million floating point operations) per image at test time. 
By deploying our method at the Princeton Starshade Testbed, we demonstrate that the neural network can be trained on simulated images and used on real images, and that it can successfully be integrated in the control system for closed-loop formation flying.
\end{abstract}

% Include a list of up to six keywords after the abstract

% Include email contact information for corresponding author
{\noindent \footnotesize\textbf{*}Corresponding author, \href{mailto:peter.melchior@princeton.edu}{peter.melchior@princeton.edu}}

\begin{spacing}{2}   % use double spacing for rest of manuscript

%%%%%%%%%%%%%%%%%%%%%%%%%%%%%%
%%%%% Introduction  %%%%%
%%%%%%%%%%%%%%%%%%%%%%%%%%%%%%
\section{Introduction}
\label{sec:intro} 

Starshades have the potential to discover and characterize the atmospheres of Earth-like exoplanets in the habitable zone of nearby stars\cite{Cash_2006, SRM, HabEx}. Their ability to achieve high contrast while maintaining high optical throughput and broad wavelength coverage make them the most promising technology to produce the first spectrum of an exo-Earth atmosphere. Starshades are currently in the technology development phase, with a targeted effort to advance critical starshade technologies (such as formation sensing and control) to Technology Readiness Level (TRL) 5\cite{S5_Plan}. A number of mission concepts have been proposed, including: the Starshade Rendezvous Mission\cite{SRM} (SRM), which pairs a 26 meter diameter starshade with the 2.4 meter diameter Nancy Grace Roman Space Telescope (NGRST; formerly known as WFIRST) separated by 27,000 km; and the Habitable Exoplanet Explorer\cite{HabEx} (HabEx), which pairs a 52 meter diameter starshade with a 4 meter diameter telescope separated by 72,000 km. In this work, we use SRM as a case study to demonstrate our methodology, but it is applicable to a starshade mission of any size.

To maintain high contrast during observations, the telescope must remain in the deep shadow cast by the starshade; typically 2 meters in diameter larger than the size of the primary mirror. The size of the shadow can be designed to be arbitrarily large, but at the expense of a larger starshade farther away, which takes more fuel and time to switch between targets. Therefore, improved formation flying allows for a more efficient starshade architecture. It has long been recognized that the most difficult aspect to starshade formation flying is accurately sensing the relative position between the starshade and telescope over thousands of kilometers separation. Knowledge of the lateral position of the starshade relative to the telescope's line of sight to the target star should be accurate to tens of centimeters in order to maintain position inside the typical $\pm1$ meter tolerance. For instance, both SRM and HabEx set a lateral position sensing requirement of 30 cm $(3\sigma)$.
%(the starshade's performance is insensitive to the axial distance to 100's of kilometers and can rely on radio ranging).
Provided accurate sensing information is available, controlling the starshade to maintain $\pm1$ meter alignment is relatively straightforward: the starshade observatory operates at the Sun-Earth Lagrange point (L2) where the dominant disturbances, gravity gradients, and solar radiation pressure are relatively benign and easily handled with linear control\cite{Flinois_2018,Flinois_2020,Palacios_2020}.

\begin{figure}
\begin{center}
\begin{tabular}{c}
  \includegraphics[width=\linewidth]{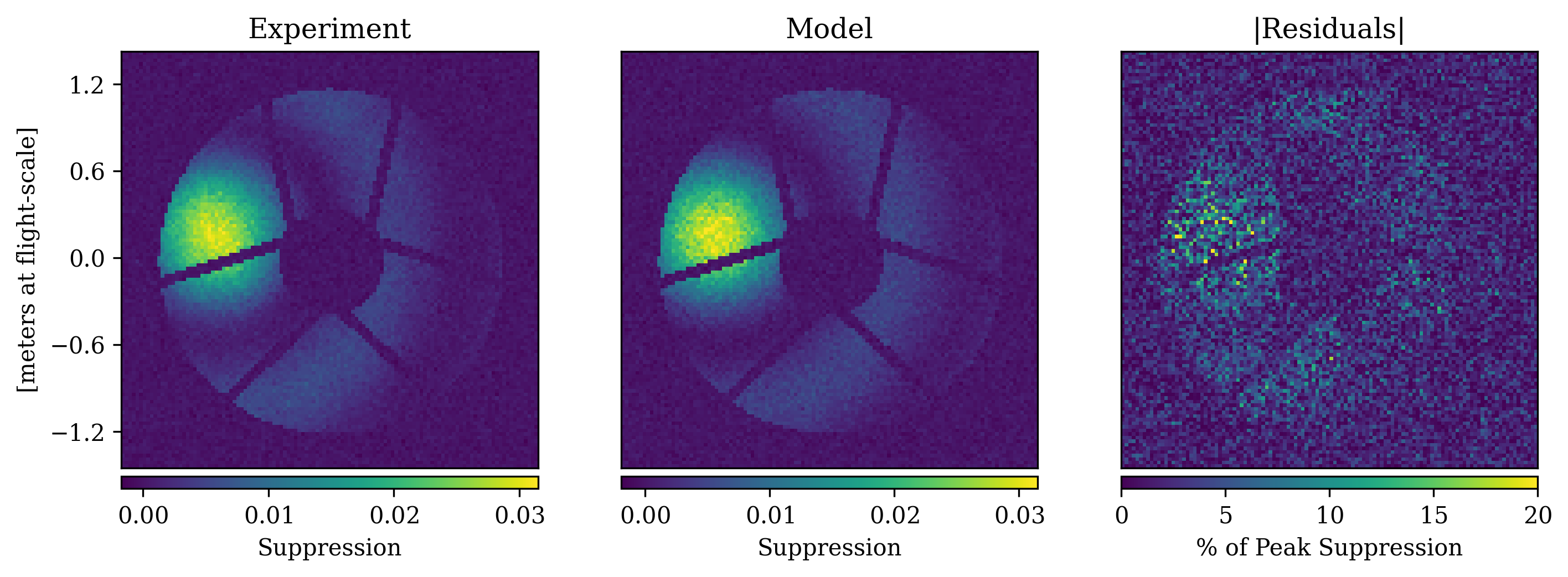}
\end{tabular}
\end{center}
\caption{\emph{Left}: experimental pupil-plane image (with Peak SNR = 5) from the Princeton Starshade Testbed. \emph{Middle}: image simulated from state-of-the-art diffraction model\cite{Harness_2021} and simulated detector noise. \emph{Right}: Residuals between experiment and model, expressed as percent of the peak suppression.\label{fig:image_comparison}} 
\end{figure}

Sensing the starshade's relative position to centimeter-level accuracy over 27,000 km is made possible by exploiting the nature of diffraction. At wavelengths outside of the starshade's designed bandpass (``out-of-band light''), its suppression performance rapidly degrades and at the center of the shadow, a bright signal of diffracted starlight reemerges as the spot of Arago (see \autoref{fig:image_comparison}). The position of this diffraction peak corresponds exactly with the position of the starshade. By imaging the diffraction pattern at the entrance pupil of the telescope, the starshade's lateral position can thus be determined to high accuracy\cite{Noecker_2007}. 

The formation sensing problem is now a matter of quickly, accurately, and reliably extracting the position of the spot of Arago from pupil plane images. Previous methods\footnote{Neural networks were also used in a very recent work\cite{Martin_2022}.} that have demonstrated sufficient accuracy in this task include an image matching algorithm\cite{Bottom_2020} and a non-linear least squares (NLLS) fit to a Bessel function model\cite{Palacios_2020}. The image matching algorithm is slow to compute, which could limit high cadence updates to the spacecraft control. It also requires a large image library, which would occupy the limited memory of the spacecraft's computer, and is limited by the finite resolution of the image library. The fit to the Bessel model is computationally intensive for state-of-the-art flight computers and requires an accurate initial guess to converge to the correct solution\cite{TDEM_2021}. 

To overcome the limitations of these previous methods, we propose a convolutional neural network (CNN) to extract the starshade position from pupil plane images. 
% Neural networks are computationally fast and only require limited onboard computing power and storage. 
In this paper we demonstrate through simulation and experiment that the CNN achieves accurate and robust position sensing with substantially reduced computing power and memory compared to previous methods.
%This work also verifies that the CNN can be trained on simulated data and still work on experimental data and in the presence of noise, which has major implications for the usefulness of this approach in a future mission.
We then make use of the CNN to produce a summary statistic for a computationally efficient simulation-based inference (SBI) technique that yields an improved estimate of the starshade position and its uncertainty.
We validate this method experimentally in the Princeton Starshade Testbed\cite{Harness_2021}, where 10$^{-10}$ contrast has been demonstrated on 1/1000$^\text{th}$ scale starshades at a flight-like Fresnel number.

The remainder of this paper is organized as follows.
In \autoref{sec:position_sensing} we pose the position sensing problem and describe the CNN architecture and training strategy. We show results on simulated and experimental data in \autoref{sec:simulation_results} and \autoref{sec:experimental_results}, respectively.
We further augment the CNN approach with a SBI technique in \autoref{sec:sbi}, and discuss our findings and future directions of this work in \autoref{sec:discussion}.

%%%%%%%%%%%%%%%%%%%%%%%%%%%%%%
%%%%% Position Sensing  %%%%%
%%%%%%%%%%%%%%%%%%%%%%%%%%%%%%
\section{Position Sensing with CNNs}
\label{sec:position_sensing}

Imaging the telescope's entrance pupil in out-of-band light shows the starshade's diffraction pattern and provides a bright signal (out-of-band light is only suppressed by a factor of 10$^{3}$, compared to 10$^{10}$ with in-band light) indicating the starshade position.
As an example, \autoref{fig:image_comparison} shows a simulated pupil image of NGRST including obstructions from the secondary mirror and supports. 

Telescopes equipped with a coronagraph typically have a camera that takes images of the pupil plane to allow low-order wavefront sensing for the coronagraph. Using pupil images therefore provides positional information without the need for additional hardware. 

\subsection{CNN Model Architecture}
\label{sec:model_arch}

\begin{figure}
\begin{center}
\begin{tabular}{c}
  \includegraphics[width=\linewidth]{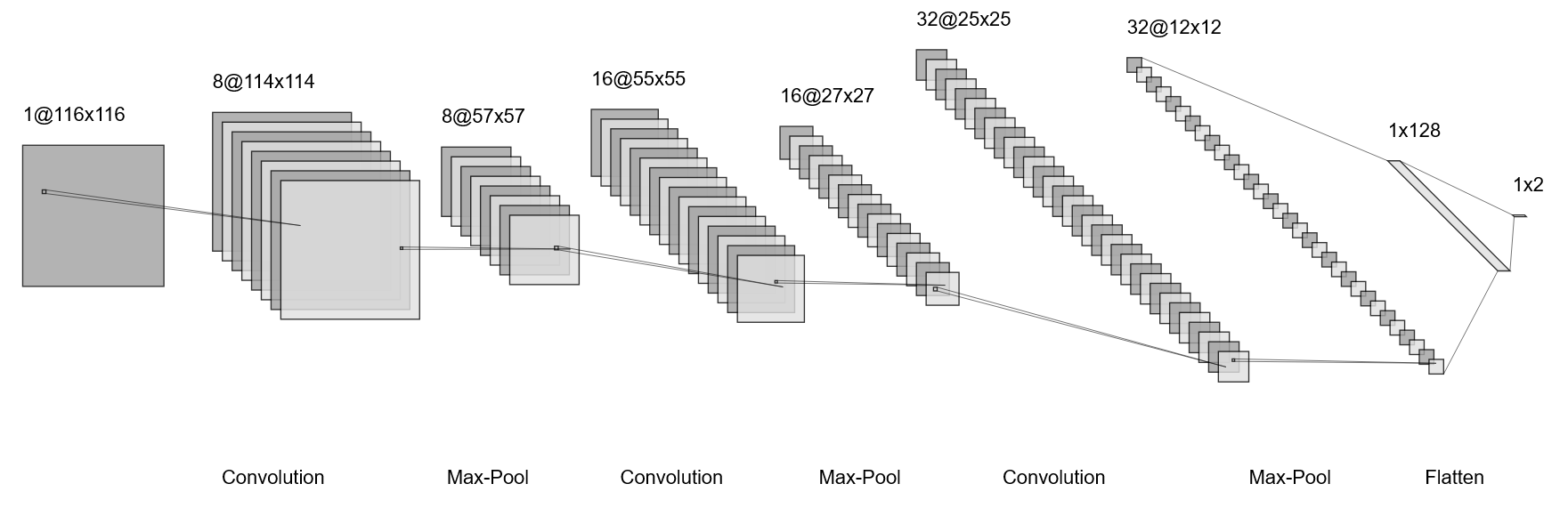}
\end{tabular}
\end{center}
\caption{CNN architecture. \label{fig:arch}
} 
\end{figure}

We decide to leverage this information by employing a CNN, a neural network architecture that is designed for image analysis and has shown excellent performance and accuracy for image classification \cite{MNIST, ImageNet_2012, ImageNet}, regression \cite{Fischer2015-ii}, and compression \cite{Cavigelli}.
We thus seek to train a CNN on the supervised regression task with simulated pupil images at true starshade positions $(x,y)$ as input and estimated positions $(x',y')$ as outputs.
To keep the computational burden low, we chose a relatively shallow CNN architecture, which is shown graphically in \autoref{fig:arch} and summarized as: 
\begin{compactitem}
    \item \textbf{Input}: 96$\times$96 pupil-plane image
    \item \textbf{Convolutional Layer}: 1 in-channel, 8 out-channels, 3$\times$3 kernel, stride 1
    \item \textbf{Max-Pooling Layer}: 2$\times$2 kernel, stride 2
    \item \textbf{ReLU Activation Function}
    \item \textbf{Convolutional Layer}: 8 in-channels, 16 out-channels, 3$\times$3 kernel, stride 1
    \item \textbf{Max-Pooling Layer}: 2$\times$2 kernel, stride 2
    \item \textbf{ReLU Activation Function}
    \item \textbf{Convolutional Layer}: 16 in-channels, 32 out-channels, 3$\times$3 kernel, stride 1
    \item \textbf{Max-Pooling Layer}: 2$\times$2 kernel, stride 2
    \item \textbf{ReLU Activation Function}
    \item \textbf{Flatten}
    \item \textbf{Fully-Connected Layer}: 4608 in-features, 128 out-features
    \item \textbf{Fully-Connected Layer}: 128 in-features, 2 out-features
    \item \textbf{Output}: estimated starshade position $(x',y')$
\end{compactitem}

\subsection{Training on Simulated Images}
\label{sec:training}

%The usefulness of any neural network is ultimately limited by how closely the training data resembles the use case. A full-scale starshade demonstration is not possible on the ground, so an experimental training set can not be generated before launch. An experimental training set could be generated on-orbit, but would use precious time and fuel. Therefore training on simulated data is the most practical way forward. 

%The limitations of training on simulated data has been well addressed\flag{cite}, but as will demonstrate in \autoref{sec:experimental_results}, this is not a limiting factor. Any starshade that flies will be extremely well characterized and its diffraction model is expected to be accurate at levels of $10^{-9}$ suppression. The simulated training data uses the same diffraction model, but with 6 orders of magnitude looser constraints, therefore the simulated diffraction pattern will be extremely accurate. The realism of the simulated training data is then determined by the calibration of the optics that form the pupil plane image and the noise properties of the detector; both of which should be well known prior to flight. The accuracy achieved in our ground-based experiment, with poorly calibrated optics and propagation through the atmosphere, demonstrates using simulated training data will not be a limiting factor. 

Simulated images are generated using the latest diffraction models\cite{Barnett_2021} that have been experimentally validated at high contrast\cite{Harness_2021}. The obstructions from NGRST's secondary mirror and its support structure are superimposed on each image and simulated photon and detector noise are added. \autoref{fig:image_comparison} shows the agreement between a simulated image and experimental image taken in the Princeton Starshade Testbed. For now, we only simulate monochromatic light to match the experimental setup.

We want the CNN to yield accurate results for different target stars, in particular in terms of brightness and spectral type.
This could be achieved by adding a conditioning variable to the last fully-connected layer, however, for this work we decided to fix the spectral type and train the CNN more robustly by varying the stellar magnitudes.
We convert the pixel values to units of suppression, which is independent of the star's brightness and exposure time. Suppression is calculated by normalizing the image by its exposure time and the photon rate of the target star without the starshade in place. The photon rate of the unblocked star is well known for all targets of interest and can be set during the pre-processing of each image. For a wider bandpass, the diffraction pattern will be slightly dependent on the spectral type of the star, but this can easily be included in the diffraction model; we defer examining the effect of a wider bandpass to a future study. 
In units of suppression, stars of different magnitudes have different signal-to-noise ratios (SNR) and different balances of photon noise to detector noise.
%(assuming a constant exposure time set by the duty cycle of the formation flying control law).
We train the CNN on images with a wide range of SNR, extending to an SNR below that expected in flight to prevent testing the CNN outside of its training conditions.
%The SNR is mainly used as a comparison between simulated images of different signal flux and exposure times; as such, 
For this work, we define the ``Peak SNR'' as the mean SNR per pixel in the FWHM of the spot of Arago. 
%When the spot of Arago is obscured by the shadow of the secondary mirror or outside the aperture, the CNN is sensing off the fainter diffraction rings, which leads to a slight reduction in performance for a given exposure time. The results in \autoref{sec:simulation_results} show that this reduction is minimal above a Peak SNR of 5. 

We define as loss function the usual L2 loss between the simulated and estimated positions. We also experimented with L1 loss, but found that the L2 loss function gave better results.
As optimizer we choose the common Adam gradient descent scheme \cite{Kingma2015-pq}. Adam is computationally efficient, uses little memory, and generally requires little hyperparameter tuning for convergence.
Finally, we use a one-cycle learning rate policy \cite{smith2018superconvergence}, which adapts the learning rate throughout the training process, changing the value at every iteration. Starting at some initial learning rate, the policy anneals the learning rate to some maximum value, then anneals it back down to some minimum value, which is typically much lower than the initial learning rate.

The training set consists of draws $(x,y)$ from a square of 3.4 m by 3.4 m around the aligned position, which covers a wider range as the $\pm$1 m formation flying requirement, again to avoid testing the CNN outside of the training conditions.
We generate 160,000 different positions, sampled uniformly from this square,
% are generated on an equally spaced 400 by 400 grid across the entire area and are randomly jittered the position in both the x and y directions, to make sure the network does not learn to predict only certain positions which are within a set grid. 
and at each position the diffraction pattern is simulated and the telescope obstructions are overlaid. For each image, the Peak SNR is randomly pulled from a uniform distribution between 0.5 and 100, and noise is then added to the image. The CNN model is trained for 30 epochs using an initial learning rate of 0.001.

%%%%%%%%%%%%%%%%%%%%%%%%%%%%%%
%%%%% Simulation Results  %%%%%
%%%%%%%%%%%%%%%%%%%%%%%%%%%%%%
\section{Simulation Results}
\label{sec:simulation_results} 

\begin{figure}
\begin{center}
\begin{tabular}{c}
  \includegraphics[width=0.9\linewidth]{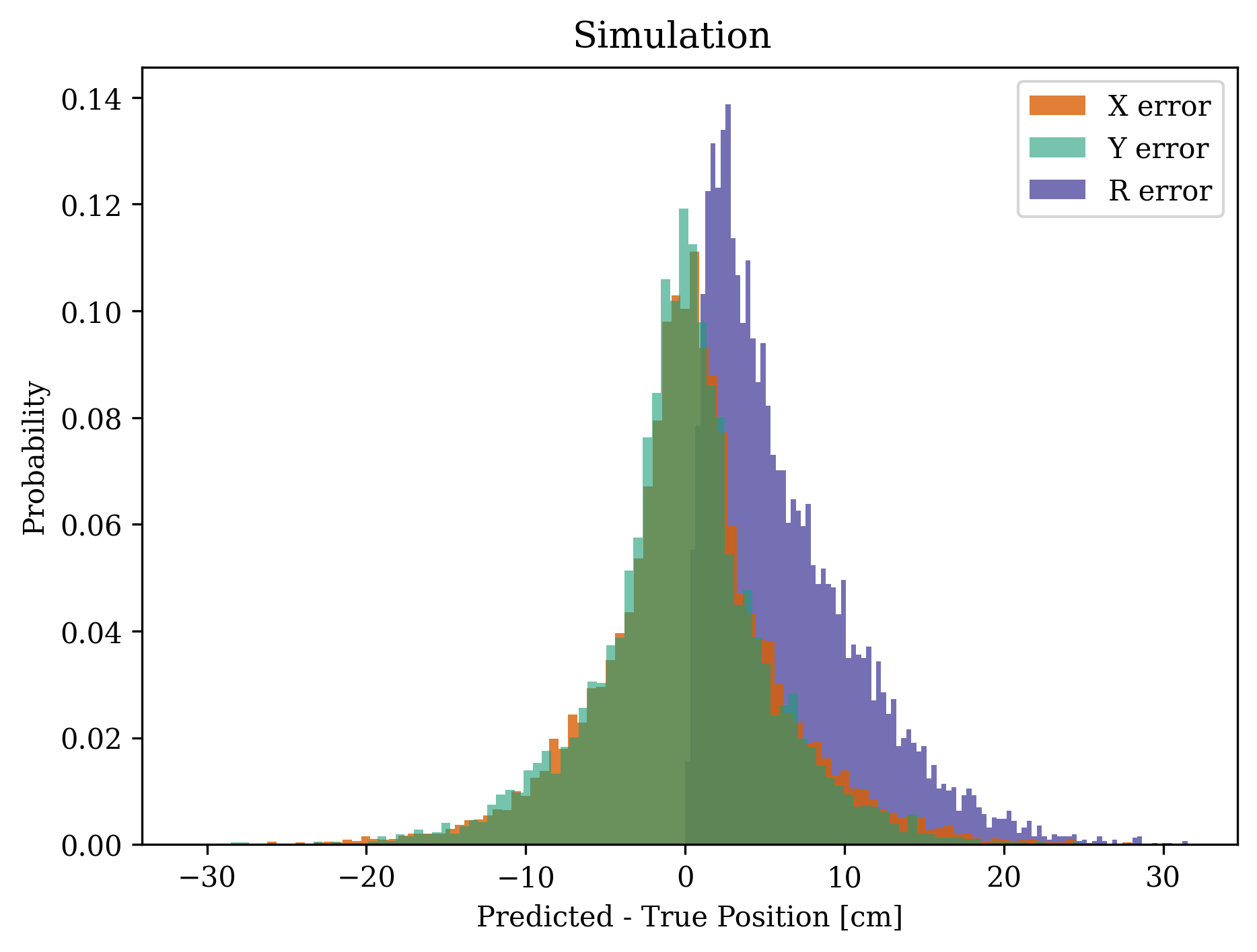}
\end{tabular}
\end{center}
\caption{{\it Simulation:} Distribution of errors in $x$, $y$ and error magnitude $R$ for the baseline simulation with Peak SNR = 5. \label{fig:base_sim_hist}} 
\end{figure}

\begin{figure}
\begin{center}
\begin{tabular}{c}
  \includegraphics[width=\linewidth]{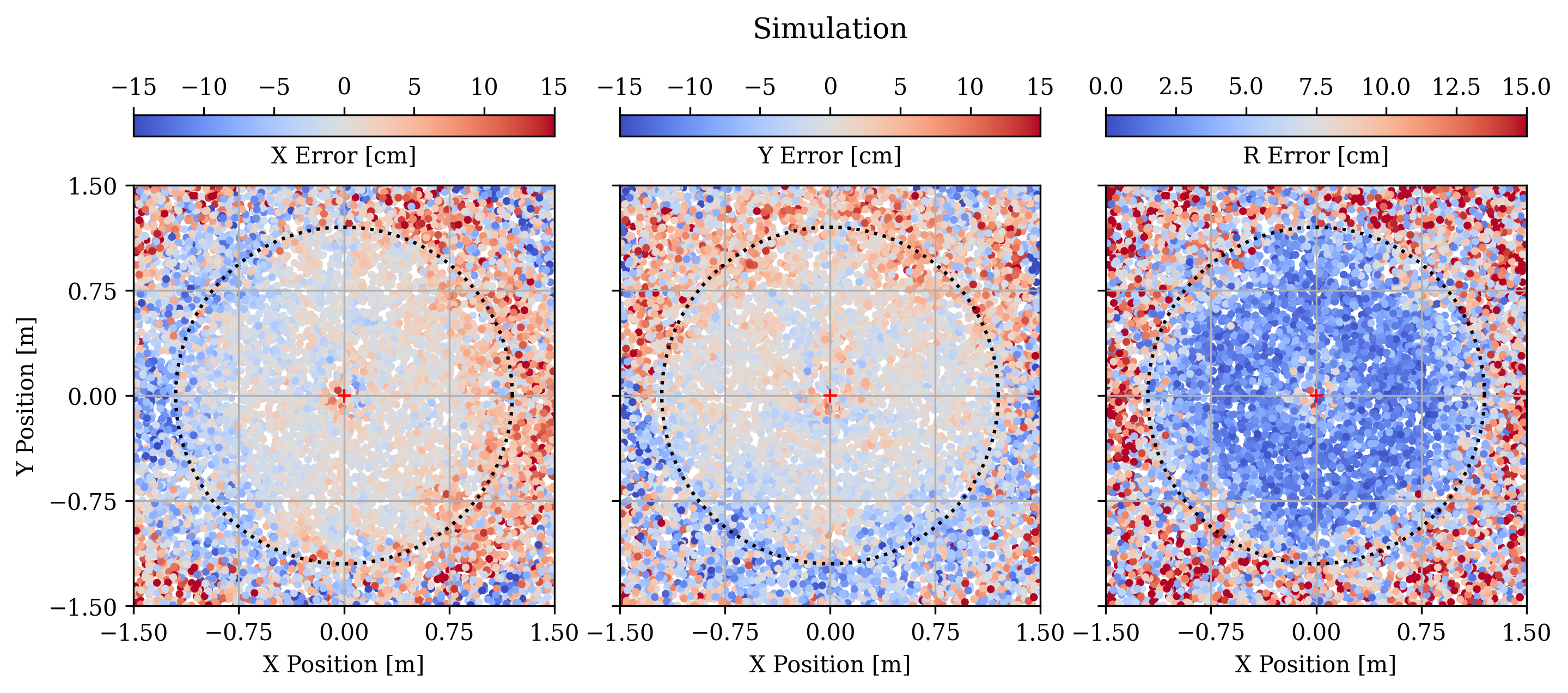}
\end{tabular}
\end{center}
\caption{{\it Simulation:} Errors in $x$, $y$ and error magnitude $R$ as a function of true starshade position for baseline simulation with Peak SNR = 5. The dotted circle marks the telescope aperture.\label{fig:base_sim_map}}
\end{figure}

We first investigate the performance of the CNN to estimate the starshade position.
We run a baseline simulation with 96$\times$96 pixels sampling the telescope's entrance aperture and a Peak SNR of 5, which corresponds to a 1 second exposure time of a 4$^\text{th}$ magnitude star with NGRST. We simulate 10,000 images with random starshade positions distributed over a 3 meter square. \autoref{fig:base_sim_hist} shows the distribution of the $(x,y)$ and magnitude $R$ position errors for the baseline simulation. The mean and standard deviation of the magnitude of the error are 6.3 cm and 4.8 cm, respectively; and 99.7\% of the errors are less than 25 cm. \autoref{fig:base_sim_map} plots these errors as a function of the starshade's position. The error increases when the spot of Arago is either obscured by the secondary mirror in the center or off the edge of the aperture. The latter situation is responsible for most of the tail of the error distribution. We thus consider 6 cm to be a conservative estimate because with an active control system the starshade should never stray past $\pm1$ meter offset during observations. If we only consider times when the offset is less than 1 meter, the mean error drops to 2.8 cm, and 99.7\% are less than 10.5 cm.

It is promising to see the good performance outside of the telescope's aperture, which suggests this method could be extended farther outside the $\pm1$ meter control box and could be used earlier in the transition between acquisition and observation guiding modes\cite{TDEM_2021}. We save the investigation into how far this method can be extended for a later date.

\begin{figure}
\begin{center}
\begin{tabular}{c}
  \includegraphics[width=\linewidth]{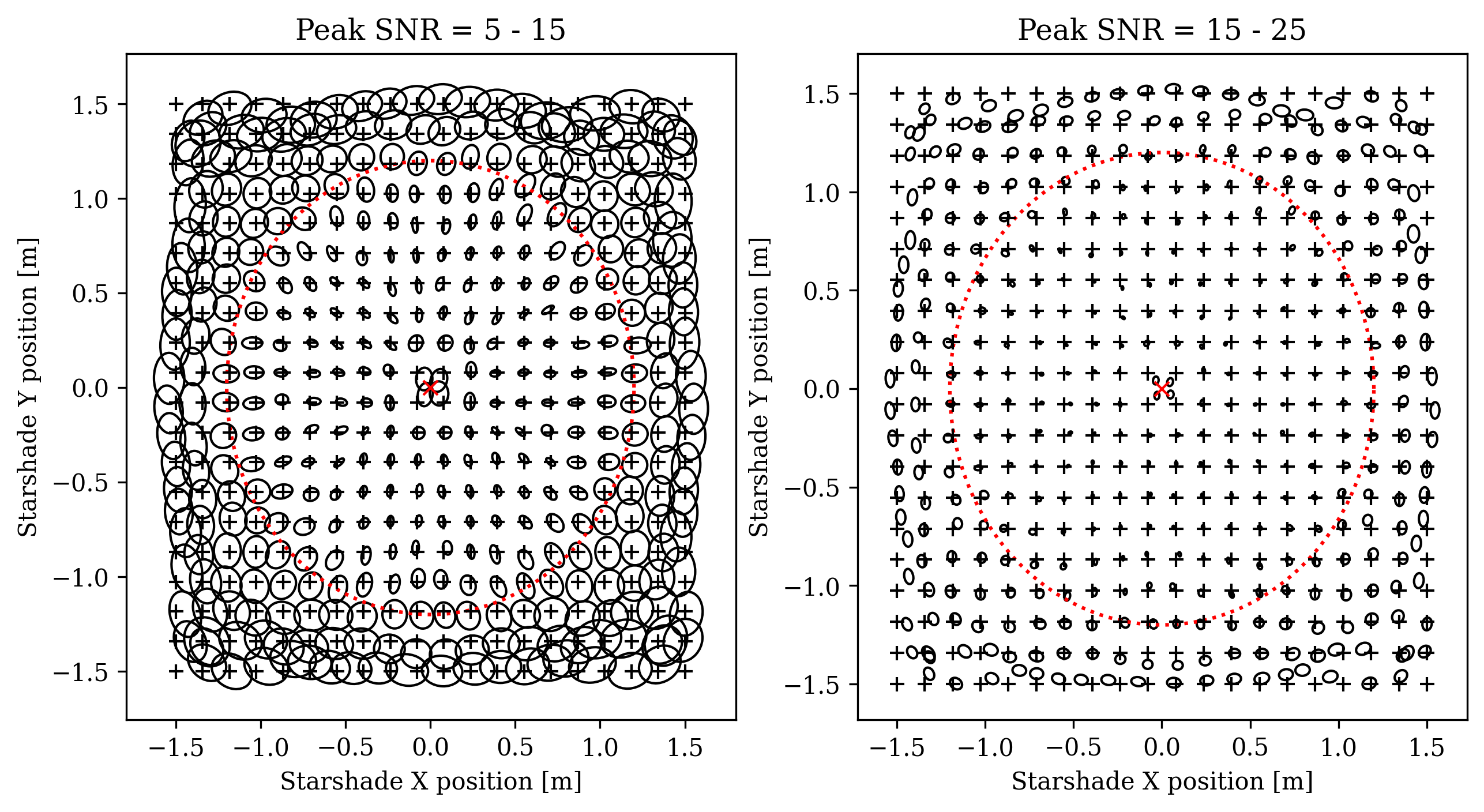}
\end{tabular}
\end{center}
\caption{Covariance ellipses as a result of 5000 simulations at each point in 20 x 20 grid (plus signs). In the {\bf left} panel the Peak SNR is between 5-15; in the {\bf right} panel the Peak SNR is between 15 - 25. The red dotted line marks the telescope aperture.\label{fig:monte_carlo}} 
\end{figure}

To examine the robustness of the CNN solution over the width of the shadow, we run a large number of simulations with the starshade positioned on a 20 x 20 grid over a 3 meter width. Each point on the grid, 5,000 simulations are run with the Peak SNR randomly pulled from a uniform distribution. \autoref{fig:monte_carlo} shows the result of this Monte Carlo simulation: the ellipses show the $3\sigma$ covariance of predicted positions centered on the mean prediction; the grid of true starshade positions are marked with black plus signs. Inside the telescope aperture, the covariance is on the order of a few centimeters and radially correlated. The left panel of \autoref{fig:monte_carlo} is the moderate SNR case with the Peak SNR drawn from 5 - 15; the right panel is the high SNR case with Peak SNR drawn from 15 - 25. There is a slight bias in the predicted positions towards the edge of the aperture, but generally this is smaller than the $3\sigma$ level. Outside the aperture, there are larger covariances and biases, with a strong inward bias at the corners. The higher Peak SNR reduces the covariances, but has little to no effect on the biases. 

\begin{figure}
\begin{center}
\begin{tabular}{c}
  \includegraphics[width=0.9\linewidth]{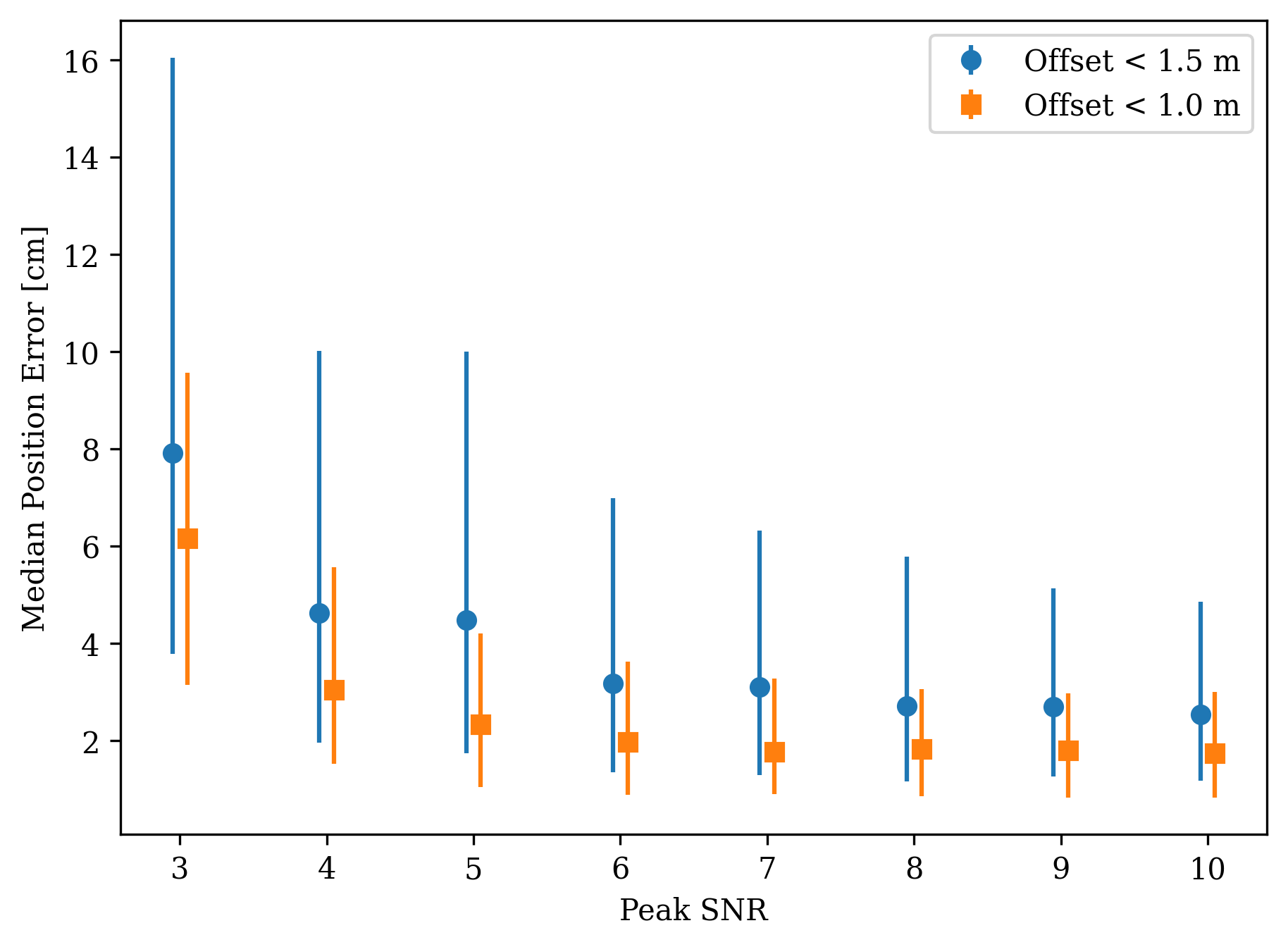}
\end{tabular}
\end{center}
\caption{Median position error as a function of Peak SNR; lower and upper error bars capture the 15.9 - 84.1 percentiles. The two curves limit the test data to offsets of a certain radius; both occur at the same SNR, but are separated for clarity.\label{fig:snr_curve}}
\end{figure}

\autoref{fig:snr_curve} shows the distribution of position errors as a function of Peak SNR. For each Peak SNR, 1000 images were simulated with randomly distributed positions; two curves are shown which limit the testset to offsets of a certain radius. We find that above a Peak SNR of 5, there is little change in the position error. Limiting to offsets of 1 meter results in a 50\% decrease in position error. 

\begin{figure}
\begin{center}
\begin{tabular}{c}
  \includegraphics[width=\linewidth]{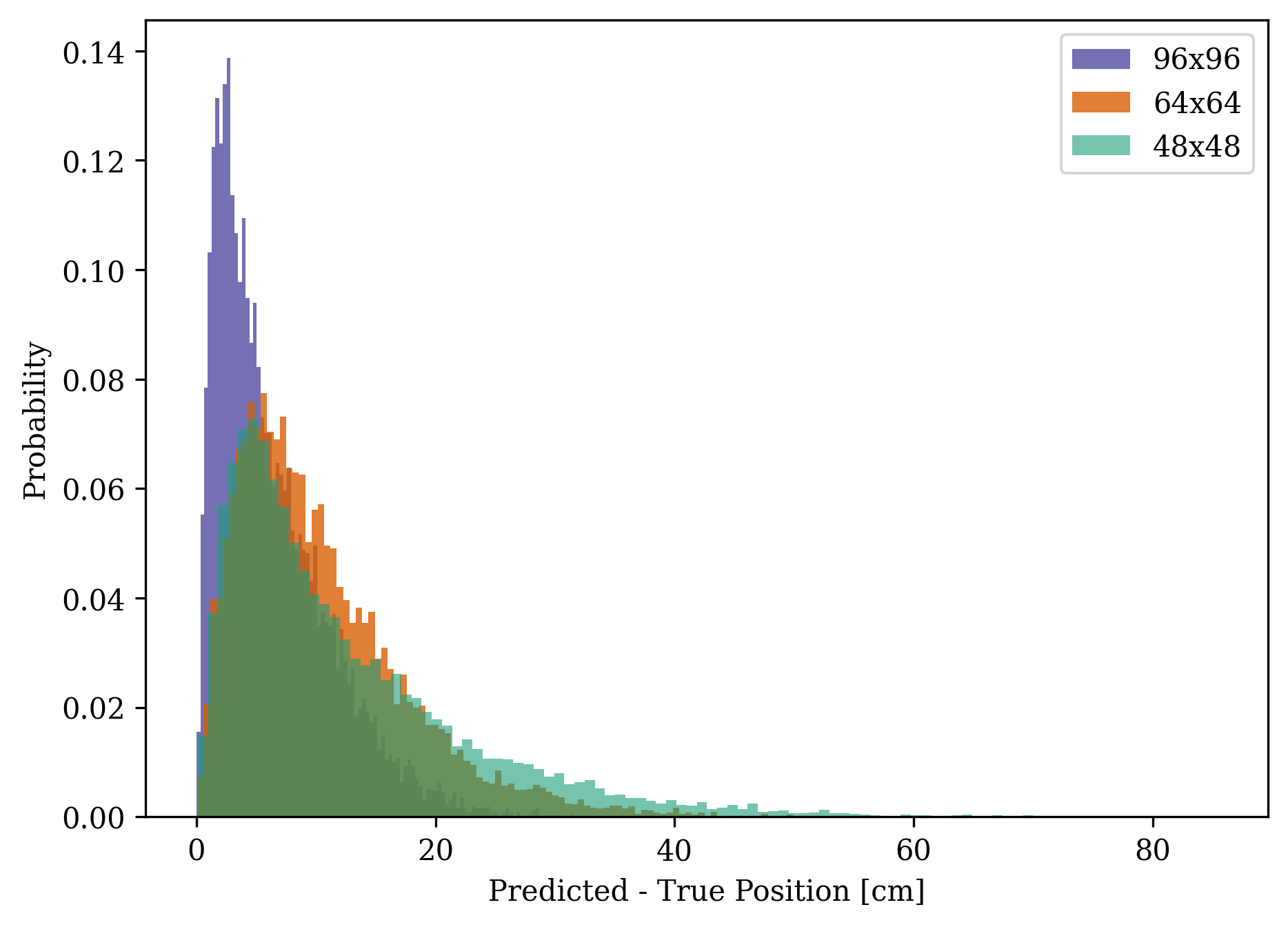}
\end{tabular}
\end{center}
\caption{Distribution of the magnitude of position errors $R$ for three image sizes. \label{fig:pixel_sampling}}
\end{figure}

Finally, we examine the trend in performance with the number of pixels sampling the pupil. \autoref{fig:pixel_sampling} shows the distribution of position errors for CNNs trained on images with sizes 48$\times$48, 64$\times$64, and 96$\times$96. We find that the position error scales approximately linearly with the number of pixels sampling the pupil (in one dimension); the mean position error for 48, 64, and 96 pixels is 12.4 cm, 10.5 cm, and 6.3 cm, respectively. As the CNN is inherently lightweight and computationally fast, the cost of a larger image size is minimal compared to the image matching and model fitting methods, and the gain in performance is worth the cost. While the choice of pupil sampling size will ultimately depend on the optical layout of the pupil imaging system and its detector electronics, the CNN benefits from a larger image size.

%%%%%%%%%%%%%%%%%%%%%%%%%%%%%%
%%%%% Experimental Results  %%%%%
%%%%%%%%%%%%%%%%%%%%%%%%%%%%%%
\section{Experimental Results}
\label{sec:experimental_results} 

The Princeton Starshade Testbed\cite{Harness_2021} was built for optical model validation experiments at high contrast levels with a flight-like Fresnel number, i.e. the parameter governing optical diffraction, but at only $1/1000$ of the scale of a real starshade.
Properties of the lab and two flight configurations are listed in \autoref{tab:lab_flight_params}.
The sub-scale starshades are lithographically etched from a silicon wafer and characterized well enough to predict the starshade performance at 10$^{-9}$ contrast. As such, the optical model should be very accurate in predicting the out-of-band diffracted light at the 10$^{-3}$ contrast level.

The testbed provides the unique opportunity to demonstrate that our approach can be trained on simulated data and then accurately be applied to experimental images.  
There are, however, a number of factors unique to the testbed that make the experimental results a conservative estimate of the on-orbit performance. First, the testbed operates in the atmosphere, and changes in the air temperature and density result in motion of the diffraction pattern and variation in the optical power incident on the starshade. Second, a low level background is observed from light scattering off aerosols as it propagates the 80 meters from laser to telescope. Third, the pupil imaging optics are off-the-shelf and not calibrated to the extent that the flight system will be, so any model mismatch seen in the lab is dominated by the pupil imaging optics and not the starshade.
Lastly, the flight missions propose using an electron-multiplying CCD, while the lab detector (properties listed in \autoref{tab:detector_noise}) is operated with a conventional amplifier and thus experiences a larger contribution from read-out noise. 

\begin{table}[!htb]
\caption{Physical parameters for the laboratory experiment and the SRM\cite{SRM} and HabEx\cite{HabEx} architectures.}
\begin{center}       
\begin{tabular}{ l | c | c | c} 
\toprule
\rule[-1ex]{0pt}{3.5ex} & {\bf Laboratory} & {\bf SRM} & {\bf HabEx}\\
\midrule 
\rule[-1ex]{0pt}{3.5ex} Telescope diameter & 2.2 mm & 2.4 m & 4.0 m\\
\rule[-1ex]{0pt}{3.5ex} Starshade diameter & 25.06 mm & 26 m & 52 m\\
\rule[-1ex]{0pt}{3.5ex} Telescope - starshade sep. & 50.0 m & 26,000 km & 76,600 km\\
\rule[-1ex]{0pt}{3.5ex} Source - starshade sep. & 27.45 m & $>3$ parsec & $>3$ parsec\\
\rule[-1ex]{0pt}{3.5ex} Formation flying tolerance & $\pm1$ mm & $\pm1$ m & $\pm1$ m\\
\rule[-1ex]{0pt}{3.5ex} Guiding bandpass & 405 nm & 425 - 552 nm & 300 - 400 nm\\ 
\rule[-1ex]{0pt}{3.5ex} Fresnel number (at $\lambda=405$ nm) & 22 & 16 & 22\\ 
\rule[-1ex]{0pt}{3.5ex} Pupil image resolution & 23 $\upmu$m/pixel & 75 mm/pixel & 125 mm/pixel \\
\bottomrule
\end{tabular}
\label{tab:lab_flight_params}
\end{center}
\end{table} 

\begin{table}[!htb]
\caption{Noise properties of testbed detector.}
\begin{center}       
\begin{tabular}{ l | c} 
\toprule
\rule[-1ex]{0pt}{3.5ex} {\bf Parameter} & {\bf Value}\\
\midrule 
\rule[-1ex]{0pt}{3.5ex} Pixel size & 13\micron$\times$13\micron\\
\rule[-1ex]{0pt}{3.5ex} Inverse gain & 0.79 e$^{-}$ / count\\
\rule[-1ex]{0pt}{3.5ex} Read noise & 4.8 e$^{-}$ / pixel / frame\\
\rule[-1ex]{0pt}{3.5ex} Dark current & 7$\times$10$^{-4}$ e$^{-}$ / pixel / s\\
\rule[-1ex]{0pt}{3.5ex} CIC noise & 2.5$\times$10$^{-3}$ e$^{-}$ / pixel / frame\\
\bottomrule
\end{tabular}
\label{tab:detector_noise}
\end{center}
\end{table}

To correct for atmosphere-induced motions in the diffraction pattern and variations in the brightness, the true position and brightness of each experimental image is solved for via the NLLS algorithm \cite{Palacios_2020}. 
% Remembering that the secondary mirror obstructions and aperture are artificially imposed after the fact, the raw images have no obstructions and see a physical aperture that is more than twice as large, providing more information for an accurate fit from the NLLS solver. 
Each image is normalized by the NLLS-solved brightness, and the NLLS-solved position is used as the true position when determining the accuracy of the CNN-DELFI position.

\begin{figure}
\begin{center}
\begin{tabular}{c}
  \includegraphics[width=0.8\linewidth]{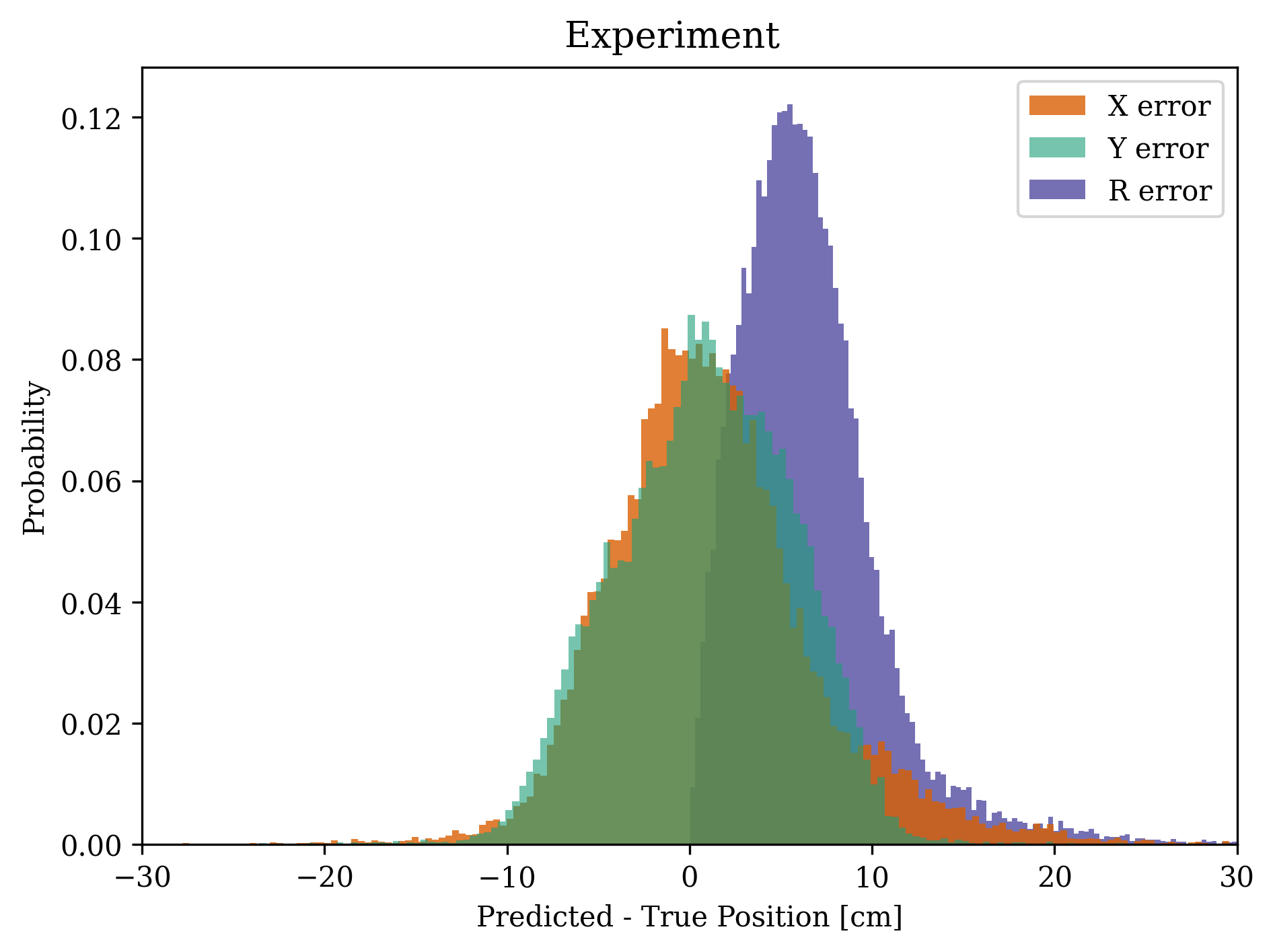}
\end{tabular}
\end{center}
\caption{{\it Experiment:} Distribution of errors in $x$, $y$ and error magnitude $R$ for Princeton Starshade Testbed experiments with Peak SNR = 5-8. \label{fig:all_lab_hist}} 
\end{figure}

\begin{figure}
\begin{center}
\begin{tabular}{c}
  \includegraphics[width=\linewidth]{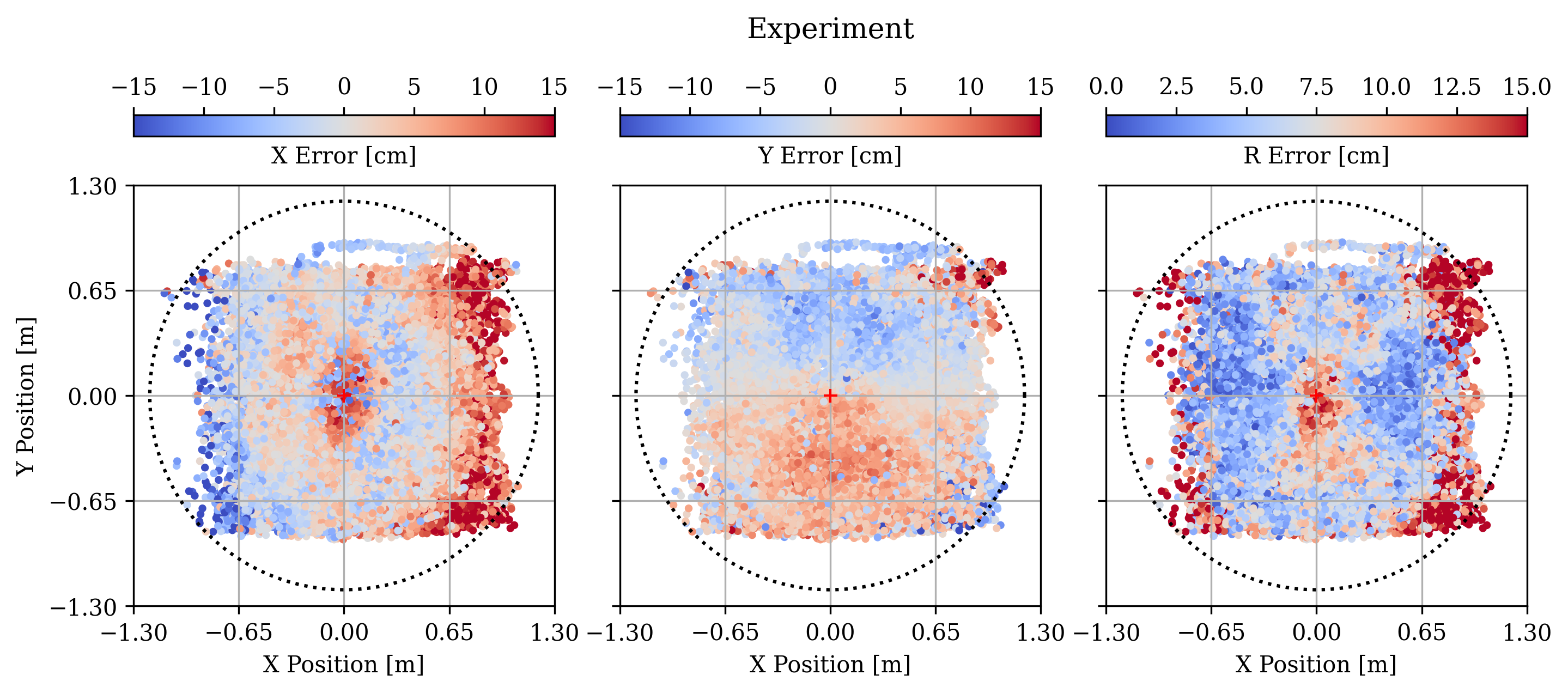}
\end{tabular}
\end{center}
\caption{{\it Experiment:} Errors in $x$, $y$ and error magnitude $R$ as a function of true starshade position  for Princeton Starshade Testbed experiments with Peak SNR = 5-8. The dotted circle marks the telescope aperture.\label{fig:all_lab_map}}
\end{figure}

\autoref{fig:all_lab_hist} shows the distribution of the X, Y, and magnitude (R) position errors for 23,700 images with Peak SNR = 5 - 8. The mean and standard deviation of the magnitude of the error are 6.5 cm and 4.0 cm, respectively, and 99.7\% of the errors are less than 25 cm. \autoref{fig:all_lab_map} plots these errors as a function of the starshade's position. The error follows the same structure as that of the simulation data, with the worst performance in the center and towards the edges, when the spot of Arago is blocked by the secondary mirror and moves off the edge of the aperture.

Sampled over the same region, the positional accuracy for experimental images is worse by about a factor of 2 compared to simulated images, likely due to the aforementioned issues with testing in the atmosphere, but the performance still exceeds our goal of 30 cm. 

\subsection{Hardware Demonstration of Closed-Loop Formation Flying}

To further validate the CNN operating in the formation flying system, we implement the CNN into the closed-loop formation-flying demonstrations presented in Refs.~\citenum{Palacios_2020, TDEM_2021}. During formation flying, the pupil camera takes an image every second and feeds it to the CNN to extract the starshade position. This position is filtered through an unscented Kalman filter (UKF) to estimate the relative position and velocity between the two spacecraft and a Linear Quadratic Regulator (LQR) controller determines the optimal control signal to minimize the starshade--telescope misalignment. 

\begin{figure}
\begin{center}
\begin{tabular}{c}
  \includegraphics[width=\linewidth]{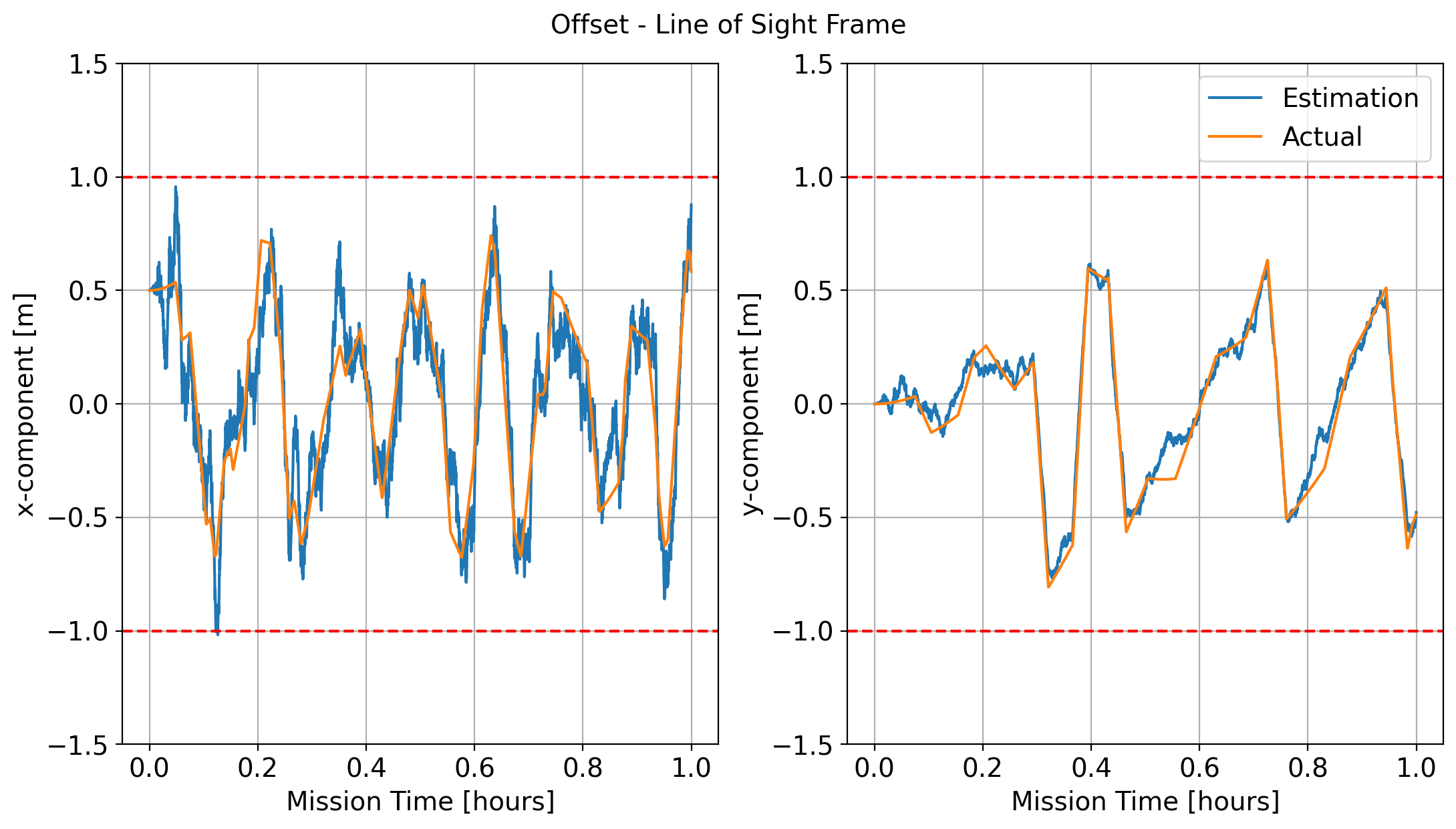}
\end{tabular}
\end{center}
\caption{{\it Experiment:} The formation-flying offset, estimated (from the Kalman filter) and actual, for the Princeton Starshade Testbed is shown in the two relevant coordinates $x$ and $y$ of the telescope’s line of sight frame. The red dashed lines mark the $\pm1$ meter requirement. \label{fig:formation_flying}}
\end{figure}

\autoref{fig:formation_flying} shows the results for an experiment with Peak SNR = 8 and with the starshade starting with a 0.5 m misalignment in the $x$-direction. Shown are the offsets between starshade and telescope, the actual and that estimated by the UKF, in the two directions of the telescope's line of sight. The estimated position tracks the truth well and the controller keeps the alignment within the $\pm 1$ meter requirement for the entire mission duration and with minimal thruster firings. The mean error from the CNN is 6.1~cm with the largest error at 23~cm. These results demonstrate the CNN can easily be integrated with existing control schemes and met the required tolerances. 

%%%%%%%%%%%%%%%%%%%%%%%%%%%%%%
%%%%% Simulation-based inference  %%%%%
%%%%%%%%%%%%%%%%%%%%%%%%%%%%%%
\section{Simulation-based Inference}
\label{sec:sbi}

The CNN produces an accurate point-estimate of the starshade position. Although it is entirely usable on its own, as we have demonstrated in the previous section, it lacks uncertainty estimates.
These would be beneficial for the control system during formation flying because the UKF could then appropriately reduce the importance of the CNN predictions in regions of large uncertainty.
We therefore couple the CNN with a subsequent calibration procedure, which provides a quantification of its associated uncertainty and yields an improved position estimate.

In keeping with the spirit of this work, we choose the lightweight method DELFI\cite{Alsing2018-rn}. In short, it amounts to fitting a Gaussian mixture model (GMM) to pairs of true $(x,y)$ and CNN-estimated $(x',y')$ positions obtained from the same simulations we introduced in \autoref{sec:position_sensing}. With a pretrained CNN, this first step is done once and the resulting GMM is stored. At test time, the GMM is evaluated at the position $(x',y')$ as reported by the CNN, which yields a probability distribution $p(x,y\mid x',y')$. The mode of this distribution is the most likely true position, while the standard deviation provides an estimate of the uncertainty.
Because of the mathematical structure of the GMM, the calculation at test time can be carried out analytically and require only matrix and vector operations in $\mathbb{R}^2$. We explain the approach in detail below.

\subsection{Density Estimation}
The first step of DELFI consists of fitting the joint distribution of true and CNN-estimated positions, i.e. $p(x,y,x',y')$.
It is useful to rephrase this problem in the typical SBI terminology, namely that we fit the joint distribution of the parameters $\btheta=(x,y)$ of the simulator and the summary statistic $\bt=(x',y')$ from the CNN.
In principle, any summary statistic that carries information about each parameter can be used, and neural networks are often trained with the specific purpose of getting a summary from the simulated data that maximize this information\cite{Charnock2018-yh}. The simplest approach is to train the network on the direct regression task, i.e. $\bt$ seeks to regress $\btheta$, as we have done in \autoref{sec:position_sensing}.

We fit the joint density with a GMM with $K=50$ components, using the python package \texttt{pygmmis}\footnote{\url{https://github.com/pmelchior/pygmmis}}\cite{Melchior2018-bk}. We repeat this process 5 times, and average the results for a GMM with $K=250$ components.
This GMM is stored for subsequent use.

\subsection{Inference}
At test time, we seek to determine the probability distribution of the parameters $\btheta$ given that we have obtained $\bt_o$ from an observed pupil image processed by the CNN.
For a single Gaussian component, this operation can be carried out analytically:

\begin{equation}
\begin{split}
\label{eq:conditional-gmm}
p(\btheta\mid \bt_o) &= \mathcal{N}(\mu_{\theta\mid t_o}, \Sigma_{\theta\mid t_o})\\
\mu_{\theta\mid t_o} &= \mu_\theta + \Sigma_{\theta, t} \Sigma_t^{-1}(\bt_o-\mu_t)\\
\Sigma_{\theta\mid t_o} &= \Sigma_\theta - \Sigma_{\theta,t}\Sigma_t^{-1}\Sigma_{\theta,t}^\top,
\end{split}
\end{equation}
where we make use of the definition of the marginal means and covariances for the joint 4-dimensional Gaussian:
\begin{equation}
\mu = \begin{bmatrix}
\mu_\theta\\
\mu_t
\end{bmatrix} \ \mathrm{and}\ 
\Sigma = \begin{bmatrix}
\Sigma_\theta & \Sigma_{\theta,t}\\
\Sigma_{\theta,t}^\top & \Sigma_t
\end{bmatrix}.
\end{equation}

For a multi-component GMM, we have to account for the relative contributions of each component at the test location $\bt_o$, which yields a new GMM in $\mathbb{R}^2$:
\begin{equation}
p(\btheta\mid \bt_o) = \sum_k \alpha_{k\mid t_o} \mathcal{N}(\mu_{\theta_k\mid t_o}, \Sigma_{\theta_k\mid t_o}),
\end{equation}
where each component mean and covariance matrix is constructed exactly in the same way as in \autoref{eq:conditional-gmm},
and the new mixture weights are
\begin{equation}
\alpha_{k\mid t_o} = \frac{\alpha_k \mathcal{N}(\bt_o\mid \mu_{t_k}, \Sigma_{t_k})}{\sum_k \alpha_k \mathcal{N}(\bt_o\mid \mu_{t_k}, \Sigma_{t_k})}.
\end{equation}
To determine an improved position estimate and its uncertainty, we can now compute the mean and covariance of this new mixture:
\begin{equation}
\begin{split}
\mu_{\theta\mid t_o} = \mathrm{mean}\left[ p(\btheta\mid \bt_o) \right] &= \sum_k \alpha_{k\mid t_o}  \mu_{\theta_k\mid t_o}\\
\Sigma_{\theta\mid t_o} = \mathrm{var}\left[ p(\btheta\mid \bt_o) \right] &= \sum_k \alpha_{k\mid t_o}  \Sigma_{\theta_k\mid t_o} + \sum_k \alpha_{k\mid t_o} (\mu_{\theta_k\mid t_o} - \mu_{\theta\mid t_o})(\mu_{\theta_k\mid t_o} - \mu_{\theta\mid t_o})^\top,
\end{split}
\end{equation}
However, a GMM is essentially guaranteed to be multimodal, so we decide to use the mode, i.e. $\mathrm{argmax}_{\btheta}\, p(\btheta\mid \bt_o)$, instead of the mean as it refers to the most common true position given the observed summary.
To speed up the process of finding the mode, we decide to simply adopt the center of the GMM component that has the highest amplitude as our improved position estimate.

\subsection{DELFI Results}

\begin{figure}
\begin{center}
\begin{tabular}{c}
  \includegraphics[width=0.8\linewidth]{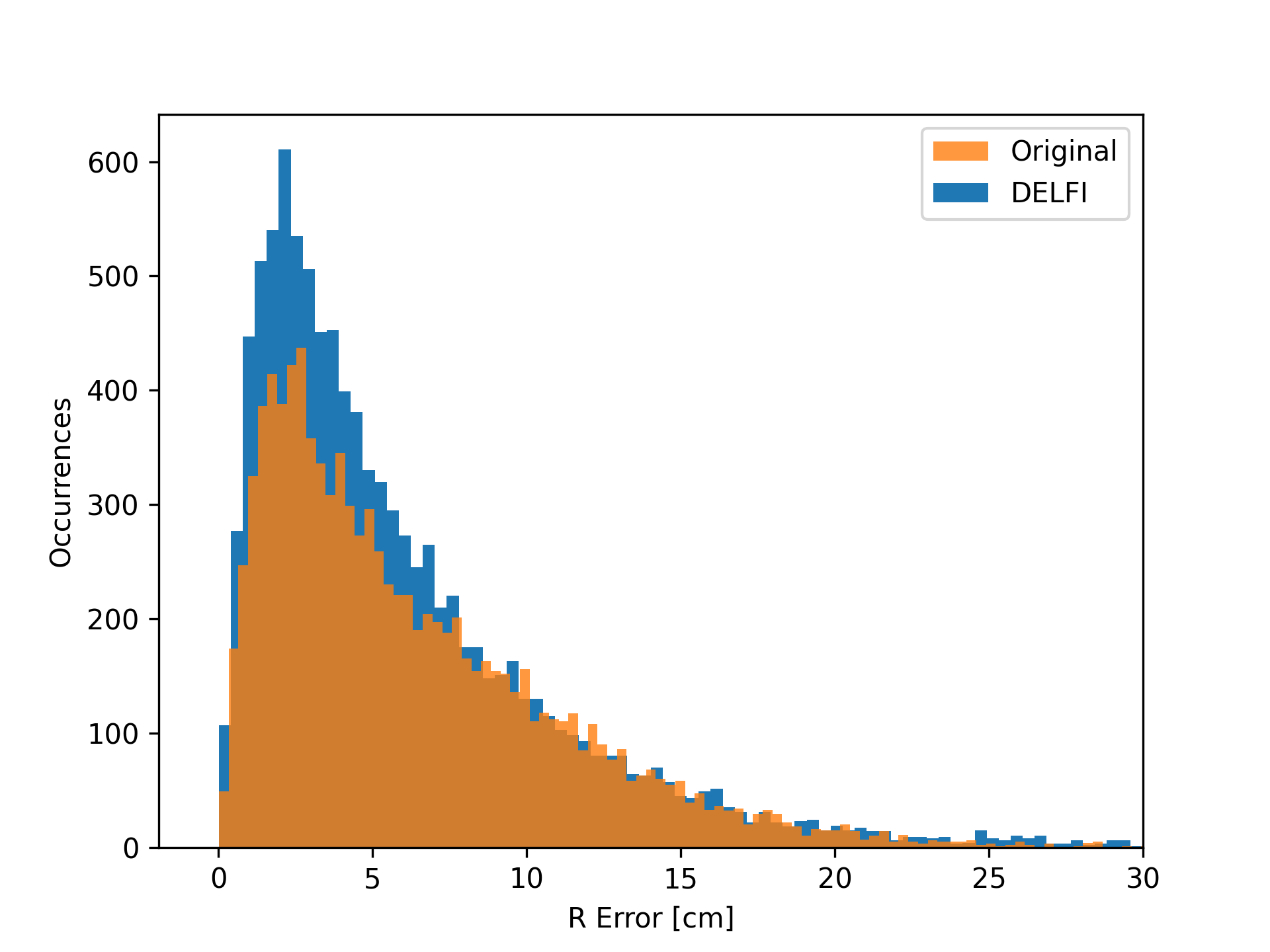}
\end{tabular}
\end{center}
\caption{\label{fig:delfi_histogram} Distribution of the error magnitude $R$ for the baseline simulation with Peak SNR = 5 for the CNN and the DELFI estimates.}
\end{figure}

We carry out the same simulations as before for the CNN test in \autoref{sec:simulation_results}.
\autoref{fig:delfi_histogram} shows a comparison of the error magnitude $R$ between the original CNN position estimates and the DELFI-corrected ones. It is evident that DELFI noticeably sharpens the peak at around $R=3$ cm by improving the estimates especially from regions where the CNN makes moderately large mistakes of $\approx$ 10 cm. 

\begin{figure}
\begin{center}
\begin{tabular}{c}
  \includegraphics[width=\linewidth]{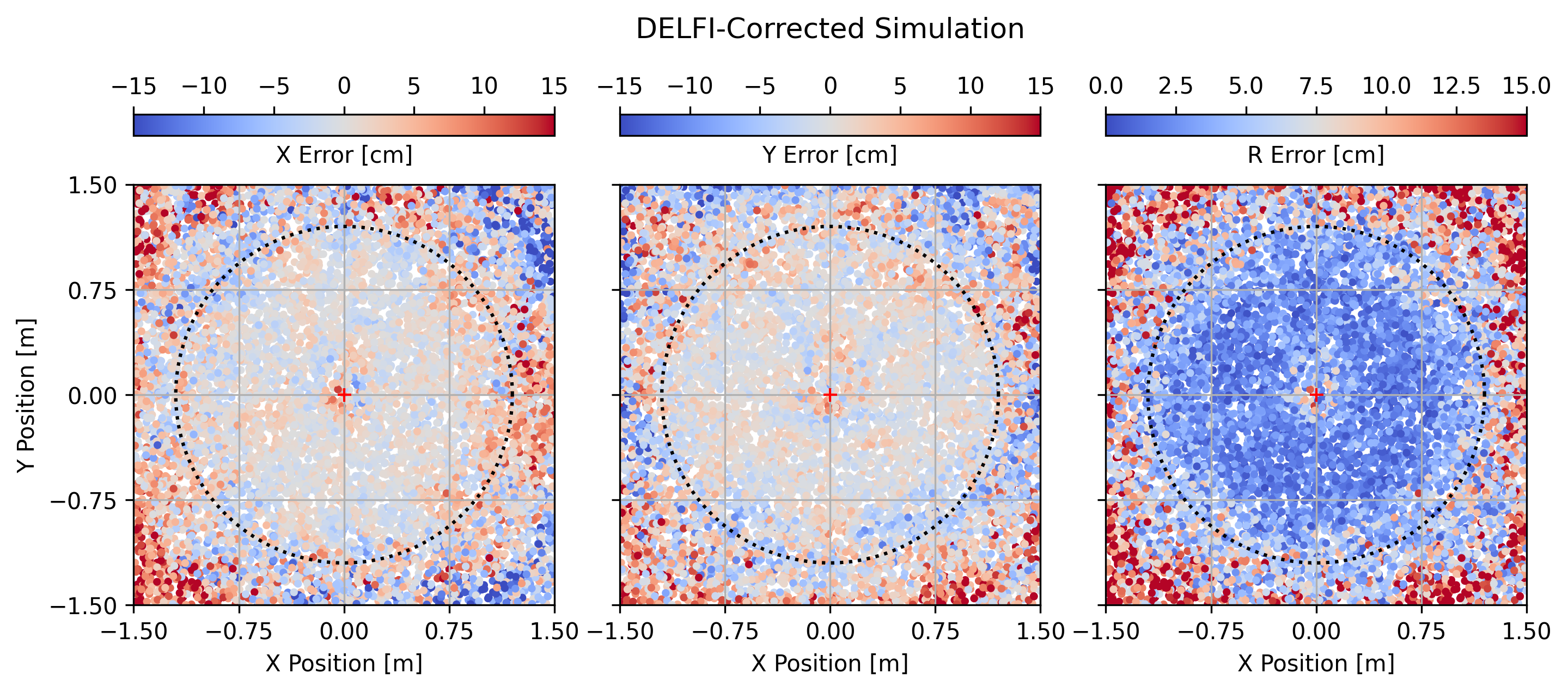}
\end{tabular}
\end{center}
\caption{\label{fig:delfi_corrected} Errors in $x, y$ and error magnitude $R$ of DELFI estimates as a function of true starshade position for baseline simulation with Peak SNR = 5. The dotted circle marks the telescope aperture.}
\end{figure}

This behavior becomes more evident in \autoref{fig:delfi_corrected}. When compared to \autoref{fig:base_sim_map}, the DELFI corrections show very little structure in the inner regions of the pupil plane and extend the range of position estimates with errors of less than $\approx$ 10 cm to well beyond the telescope aperture radius.

\begin{figure}
\begin{center}
\begin{tabular}{c}
  \includegraphics[width=\linewidth]{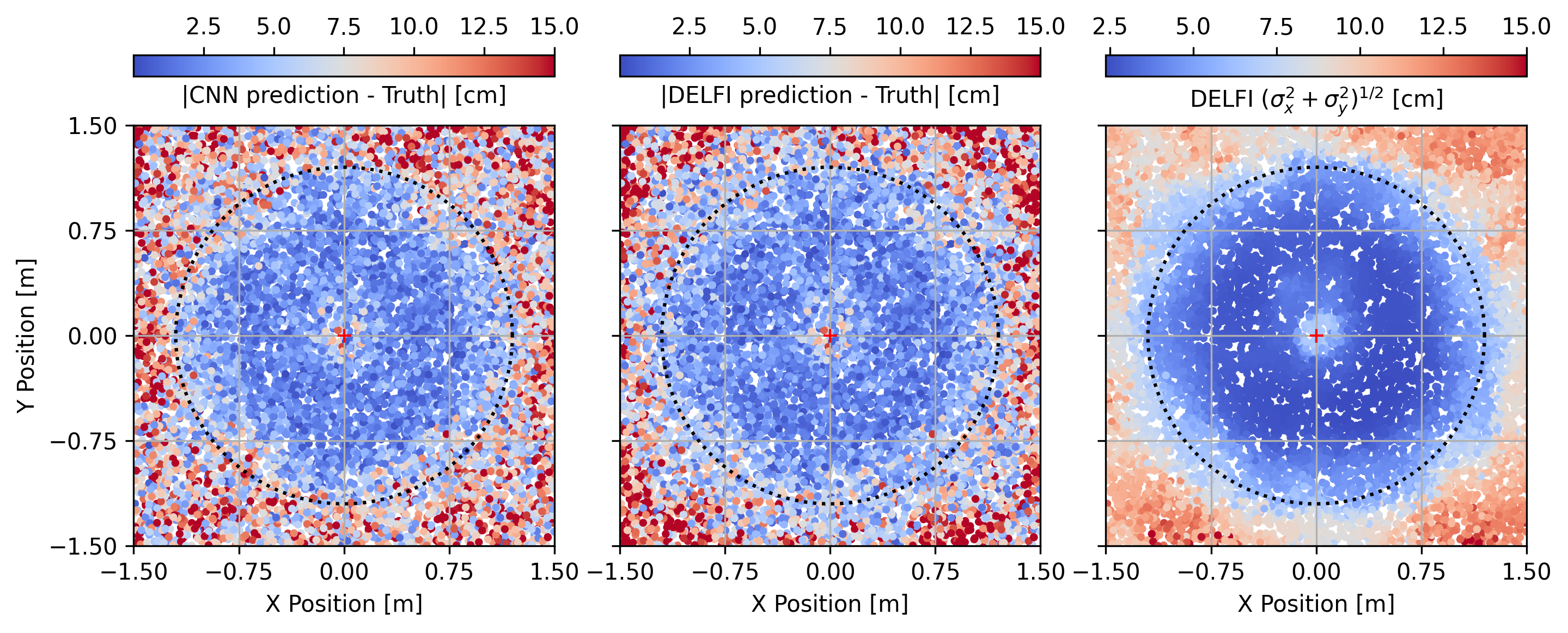}
\end{tabular}
\end{center}
\caption{\label{fig:delfi_results} Error magnitude $R$ as a function of true starshade position for the baseline simulation with Peak SNR = 5  for the CNN estimates (\emph{left}) and the DELFI estimates (\emph{middle}). The \emph{right} panel shows the predicted uncertainty of $R$ from DELFI and matches closely the actual estimation accuracy. The dotted circle marks the telescope aperture.}
\end{figure}

As final demonstrate of the calibration procedure, \autoref{fig:delfi_results} show the error magnitude as function of position offset for the CNN and for DELFI. The right panel shows the predicted magnitude uncertainty obtained from DELFI, which matches very well with the accuracy of its own estimates in the middle panel. The areas of large uncertainty remain those where the information from the pupil image is physically limited, namely when the spot of Arago is blocked either by the secondary mirror in the center or the outer perimeter of the telescope pupil.
We can therefore conclude that DELFI provides well-calibrated position estimates and their uncertainties for further use by the control system for formation flying.

\subsection{Impact on formation flying}
To examine the impact of the DELFI provided uncertainty, we ran a suite of formation flying simulations\cite{Palacios_2020} with the position uncertainty given either by the DELFI estimate, or by assuming a constant uncertainty of 2~cm. We created 100 random initial starting positions within a 1 meter offset and ran a simulated 1 hour mission for each uncertainty case. For the DELFI provided uncertainty, the average fuel consumption was 3\% lower and the error in the UKF estimated position was 8\% lower. With this quick check, we conclude that DELFI provides a better estimate of the measurement uncertainty that leads to a more efficient state estimation. While the effect size shown here is small, it is more important to note that this method provides a reasonable estimate of the uncertainty, and it is possible for future control schemes to utilize this information to its full extent.

%%%%%%%%%%%%%%%%%%%%%%%%%%%%%%
%%%%% Discussion  %%%%%
%%%%%%%%%%%%%%%%%%%%%%%%%%%%%%
\section{Conclusion and Outlook}
\label{sec:discussion} 

The results of \autoref{sec:simulation_results} and \autoref{sec:experimental_results} show that the CNN alone can easily meet the position estimation requirement of 30 cm set in mission designs\cite{SRM, HabEx}. For the entirety of the observation phase, the control law should not allow the misalignment to exceed 1 meter, therefore we focus on results where the starshade position is less than 1 meter. When doing so, the 99.7 percentile position error for the baseline simulation is 10.5 cm, and for the experimental results is 21.5 cm.\footnote{The experimental results should be viewed as a conservative estimate of the performance as the mismatch between experimental and simulated training data is dominated by the effects of the atmosphere and uncharacterized pupil imaging optics. For an actual mission, atmospheric issues will not be present in space, and the imaging system will be better characterized on the ground and after launch for an accurate model of the instrument response.}
These results meet the 30 cm requirement that defines the formation sensing milestone for TRL 5\cite{S5_Plan}.
As we demonstrate in \autoref{sec:sbi}, augmenting the CNN with our simulation-based calibration method DELFI further improves the estimation accuracy and pushed out typical errors of less than 10 cm to well beyond the telescope pupil.

Our results compare favorably to alternative approaches with image matching\cite{Bottom_2020} and NLLS\cite{Palacios_2020} algorithms, all of which have been shown to satisfy accuracy requirements. The most significant advantages of our approach is that it is lightweight in memory and fast to compute.
The CNN trained on images of 96$\times$96 pixels has $4\times10^5$ parameters and takes up 1.6~MB in disk space. A CNN trained on 32$\times$32 images only needs 92~kB of disk space, 500$\times$ less than the 46~MB required for the image library (32$\times$32) of the matching algorithm\cite{Bottom_2020}. A DELFI model with $K=250$ Gaussian components also only occupies 42~kB.
The CNN needs 5.3 MFLOPs (million floating point operations) to process a 96$\times$96 image; 0.36 MFLOPs to process a 32$\times$32 image. The computational cost of DELFI is entirely negligible. For the image library method, Ref.~\citenum{Bottom_2020} reports 70 MFLOPs to search the whole library, but suggest that a reduced grid search around an initial guess could reduce that to 0.5 MFLOPs. The NLLS algorithm involves computing \autoref{eq:bessel_model} and its analytical Jacobian at each pixel once per iteration; with a good initial guess, the Levenberg-Marquardt solution typically converges within 15 iterations. The NLLS computation time is dominated by the calculation of the Bessel function and its derivative ($J_0$, $J_1$), which we estimate to require $3BN$ FLOPs each ($B$ is the number of terms kept in the Bessel function series expansion and $N$ is the total number of pixels). Therefore, the NLLS takes 14 MFLOPs to process a 96$\times$96 image; 1.6 MFLOPs to process a 32$\times$32 image. 

Our method thus substantially reduces storage space and computation time over the image matching algorithm. It also produces a continuous position estimate, whereas the image matching results are discrete due to the finite offset sampling in the image library. Our method also outperforms the NLLS algorithm, though to a lesser degree, and provides the following extra benefits: 1) unlike NLLS, it does not require a valid initial guess to properly converge, and 2) the Bessel function model used by the NNLS is only valid out to an offset of 1.5 meters. Future work can investigate whether additional training of the CNN model and DELFI renders it effective for larger offsets.

One important aspect of practical applicability we did not address in this work is related to variations in the stellar spectrum. Due to the monochromatic laser in the experimental testbed, this study was limited to light of a single wavelength and therefore does not capture the effect of the target star's spectral type $t$.
While we expect that CNN estimates to only weakly depend on $t$,%
\footnote{Over the 100 nm wide guiding bandpass, the diffraction pattern is well-approximated by the Bessel function model:
\begin{equation}
    I(s) = J^2_0\left(\frac{2\pi R s}{\lambda z}\right) \,,
    \label{eq:bessel_model}
\end{equation}
where $I$ is the intensity as a function radial coordinate $s$, $R$ is the starshade radius, $z$ is the telescope--starshade separation, and $\lambda$ is the wavelength of light. Inverting \autoref{eq:bessel_model}, the width of the spot of Arago scales linearly with wavelength (as $\lambda z /R$). To first order, and due to the stellar flux increasing with wavelength on the Wien's tail for solar-type stars, the diffraction pattern will be dominated by the spot of Arago broadened by the longest wavelength in the bandpass. As the CNN is trained on units of suppression and images are normalized by the unblocked star's expected flux, the CNN is expected to be weakly dependent on spectral type $t$.}%
future works should be able to improve the accuracy for broadband images by training the CNN on image with a range of spectral types. For predictions it can then either implicitly marginalize over an unknown spectral type or, even better, condition its estimates on the known spectral type $t$ of the target star. The latter option can be achieved by adding the spectral type as an extra variable to the last fully connected layer of the CNN, at a minor increase of required memory. Our calibration method DELFI can also take the spectral type into account by evaluating $p(x,y\mid x', y', t)$, i.e. by modestly increasing the task of density estimation from a four- to a five-dimensional space.

We therefore conclude that position sensing with CNN and DELFI is a viable, accurate, and computationally efficient option for formation flying of future Starshade missions. To help with the adoption of our method, we publicly release the code at \url{https://github.com/astro-data-lab/starshade-xy}.

%%%%%%%%%%%%%%%%%%%%%%%%%%%%%%
%%%%% End material  %%%%%
%%%%%%%%%%%%%%%%%%%%%%%%%%%%%%

\subsection*{Disclosures}
The authors have no disclosures to report.

\acknowledgments 
Part of this work was supported by NASA Technology Development for Exoplanet Missions (TDEM) award: NNX15AD31G. AH is supported by the Jet Propulsion Laboratory, California Institute of Technology under a contract with the National Aeronautics and Space Administration. 

%%%%% References %%%%%

\bibliography{references}   % bibliography data in references.bib
\bibliographystyle{spiejour}   % makes bibtex use spiejour.bst

% %%%%% Biographies of authors %%%%%
% \newpage
% \vspace{2ex}\noindent\textbf{Anthony Harness} is an Associate Research Scholar in the Mechanical and Aerospace Engineering Department at Princeton University. He received his Ph.D. in Astrophysics in 2016 from the University of Colorado Boulder. He currently leads the experiments at Princeton validating starshade optical technologies. 

% \listoffigures
% \listoftables

\end{spacing}
\end{document}